\documentclass[aps,superscriptaddress,twocolumn,nofootinbib]{revtex4}



\usepackage{natbib}

\usepackage{graphicx}
\usepackage{dcolumn}
\begin{document}


\title{On the origin of power law tails in price fluctuations}




\author{J. Doyne Farmer}
\affiliation{Santa Fe Institute, 1399 Hyde Park Road, Santa Fe, NM 87501}

\author{Fabrizio Lillo}
\affiliation{Santa Fe Institute, 1399 Hyde Park Road, Santa Fe, NM 87501}
\affiliation{Istituto Nazionale per la Fisica della Materia, Unit\`a di Palermo, Italy}


\date{\today}

\begin{abstract}
In a recent Nature paper, Gabaix et al. \cite{Gabaix03} presented a
testable theory to explain the power law tail of price fluctuations.
The main points of their theory are that volume fluctuations, which
have a power law tail with exponent roughly $-1.5$, are modulated by
the average market impact function, which describes the response of
prices to transactions.  They argue that the average market impact
function follows a square root law, which gives power law tails for
prices with exponent roughly $-3$. We demonstrate that the long-memory
nature of order flow invalidates their statistical analysis of market
impact, and present a more careful analysis that properly takes this
into account.  This makes it clear that the functional form of the
average market impact function varies from market to market, and in
some cases from stock to stock.  In fact, for both the London Stock
Exchange and the New York Stock Exchange the average market impact
function grows much slower than a square root law; this implies that
the exponent for price fluctuations predicted by modulations of volume
fluctuations is much too big.  We find that for LSE stocks traded in
the electronic market the distribution of transaction volumes does not
even have a power law tail.  This makes it clear that volume
fluctuations do not determine the power law tail of price returns.
\end{abstract}


\maketitle

Gabaix {\it et al.} \cite{Gabaix03} have recently proposed a testable
theory for the origin of power law tails in price fluctuations.  In
essence, their proposal is that they are driven by fluctuations in the
volume of transactions, modulated by a deterministic market impact
function.  More specifically, they argue that the distribution of
large trade sizes scales as $P(V>x) \sim x^{-\gamma}$, where $V$ is
the volume of the trade and $\gamma \approx 3/2$.  Based on the
assumption that agents are profit optimizers, they argue that the
average market impact function\footnote{One should more properly think
  of the market impact as a response to the order initiating the
  trade.  That is, in every transaction there is a just-arrived order
  that causes the trade to happen, and this order tends to alter the
  best quoted price in the direction of the trade, e.g. a buy order
  tends to drive the price up, and a sell order tends to drive it
  down.} is a deterministic function of the form $r = k V^{\beta}$,
where $r$ is the the change in the logarithm of price resulting from a
transaction of volume $V$, $k$ is a constant, and $\beta = 1/2$.  This
implies that large price returns $r$ have a power law distribution
with exponent $\alpha = \gamma/\beta \approx 3$.  They argue that
their theory is consistent with the data, even though their hypothesis
about market impact appears to contradict several other previous
studies \cite{Plerou02,Lillo03,Bouchaud03} in the same markets they
study (the New York and Paris Stock Exchanges).

\section{Problems with the test of Gabaix et al.}

Gabaix et al. \cite{Gabaix03} present statistical evidence that
appears to show that the NYSE and Paris data are consistent with the
hypothesis that the average market impact follows a square root law.
In this section we show that their test may have problems in
circumstances (such as those of the real data) in which orders have
long-memory properties.  This weakens their test, so that
it lacks the power to reject reasonable alternative hypotheses and
may give misleading results.

Their method to test the hypothesis of square root price impact is to
investigate $E[r^2|V]$ over a given time interval, e.g. $15$ minutes,
where $r$ is the price shift and $V=\sum_{i=1}^{M} V_i$ is the sum of
the volumes of the $M$ transactions occurring in that time interval.
They have chosen to analyze $r^2$ rather than $r$ because of its
properties under time aggregation.  To see why this might be useful,
assume the return due to each transaction $i$ is of the form $r_i = k
\epsilon_i V_i^{\beta} + u_i$, where $u$ is an IID noise process that
is uncorrelated with $V_i$, and $\epsilon_i$ is the sign of the
transaction.  The squared return for the interval is then of the form
\begin{equation}
r^2 = \sum_{i=1,j=1}^{M} (k\epsilon_i V_i^\beta + u_i)
(k\epsilon_j V_j^\beta + u_j)
\end{equation}
Under the assumption that $V_i$, $V_j$, $\epsilon_i$, and $\epsilon_j$
are all uncorrelated, when $\beta = 1/2$ it is easy to show that
$E[r^2|V]=a + b~V$, where $a$ and $b$ are constants.

\begin{figure}[ptb]
 \begin{center} 
 \vspace{-.5in}
 \includegraphics[scale=0.3]{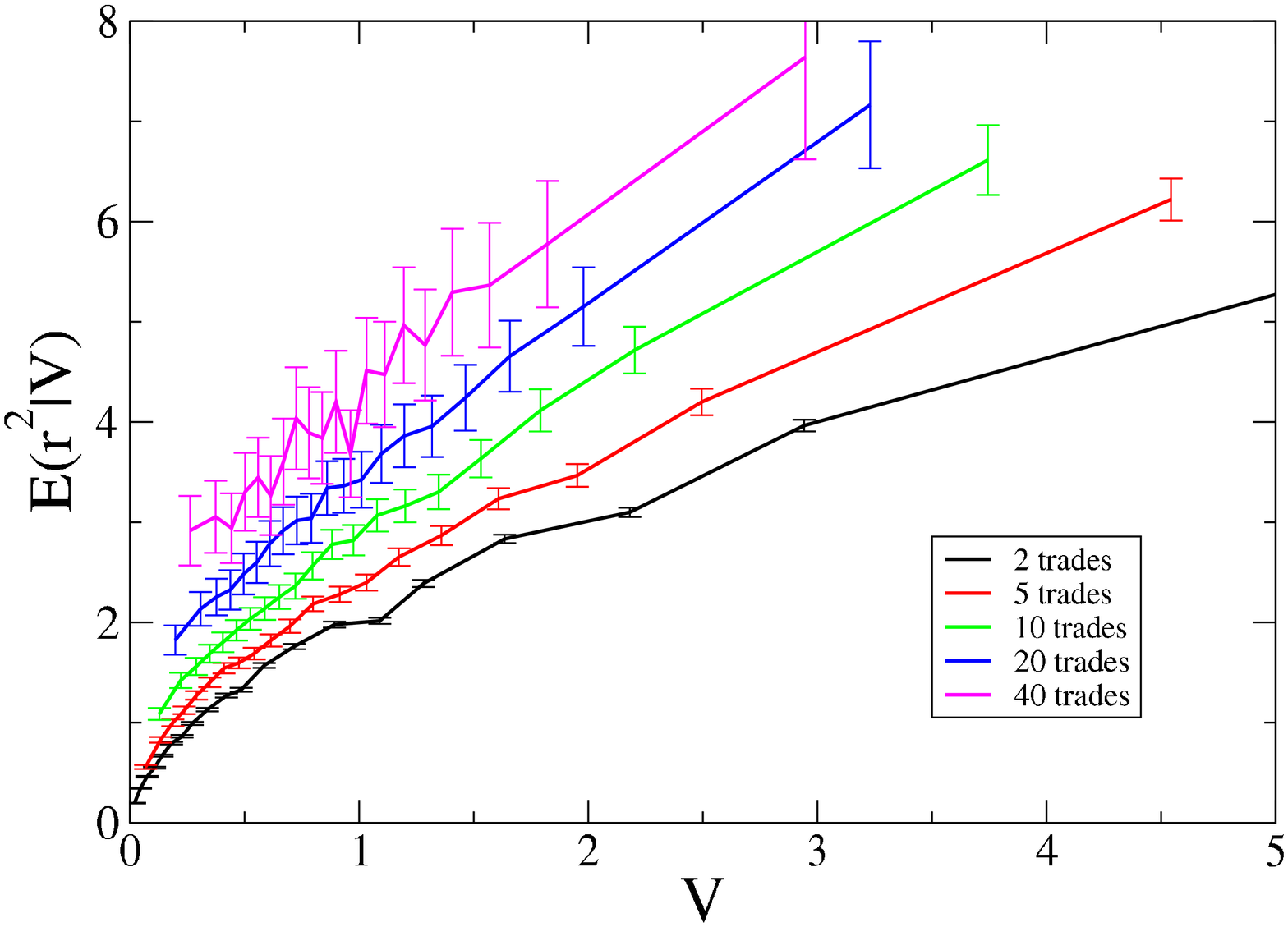}
 \vspace{-.2in}
 \includegraphics[scale=0.3]{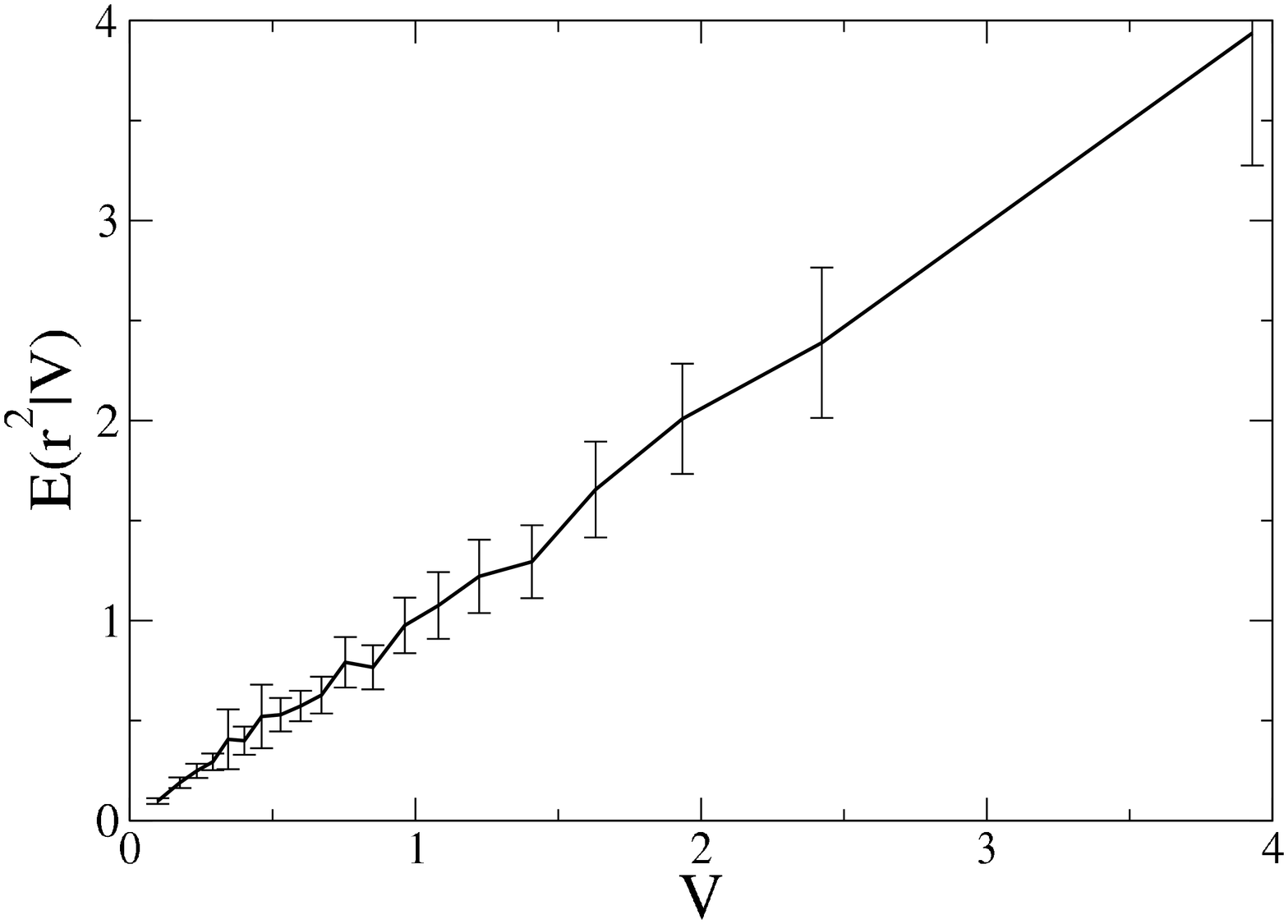}
 \vspace{-.1in}
 \includegraphics[scale=0.3]{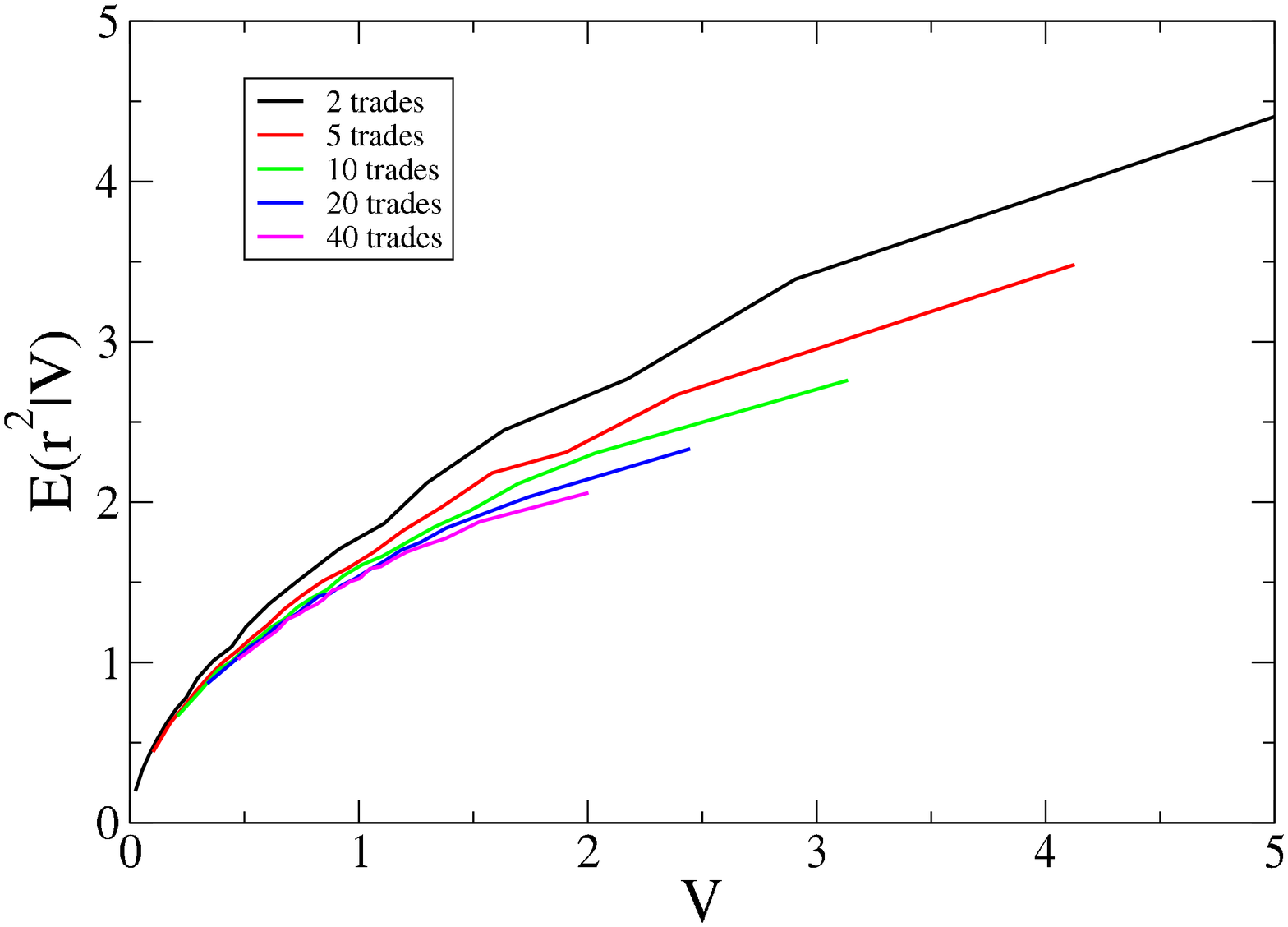} 
 \caption{A demonstration that the statistical test of Gabaix et
   al.~\cite{Gabaix03} fails due to the strong autocorrelations in
   real data.  The expected value of the squared price return,
   $E[r^2|V]$, is plotted as a function of total transaction size
   $V=\sum_{i=1}^M V_i$, where $V_i$ is the size of transaction $i$.
   Each transaction causes a simulated market impact of the form $r_i
   = k \epsilon_i V_i^{\beta}$, to generate total return $r =
   \sum_{i=1}^M r_i$.  The transaction series $V_i$ and $\epsilon_i$
   are from the real data from the electronic market for the British
   stock Vodafone, and contain roughly $10^6$ events.  The error bars
   are the $95\%$ confidence intervals computed following the
   procedure specified by Gabaix et al.  (a) shows the results for a
   fixed number of transactions, with $M$ varying from $2$ to $40$;
   the curves are in ascending order of $M$; (b) is the same using a
   fixed time interval of $15$ minutes, with variable $M$; and (c) is
   the same as (a) with the order of the transactions randomly
   shuffled.  For (a) and (b) we see straight lines for large $M$,
   indicating that the test is passed, even though by construction the
   market impact does not follow the $r\sim V^{0.5}$ hypothesis,
   whereas for the shuffled data the test quite clearly shows us that
   the hypothesis is false.}
\label{expRsq}
\end{center}
\end{figure}

The problem is that for the real data $V_i$, $V_j$, $\epsilon_i$, and
$\epsilon_j$ are strongly correlated, and indeed, the sequence of
signs $\epsilon_i$ is a long-memory process
\cite{Bouchaud03b,Lillo03b}.  To demonstrate the gravity of this
problem, we use real transactions $V_i$, but introduce an artificial
and deterministic market impact function of the form $r_i = k
V_i^{\beta}$ with $\beta\neq 0.5$.  We first fix the number of
transactions, and then repeat the same procedure using a fixed time
period.  We examine blocks of trades with $M$ transactions,
$\lbrace\epsilon_i, V_i \rbrace$, $i=1,...,M$, where $\epsilon_i=+1$
($-1$) for buyer (seller) initiated trades and $V_i$ is the volume of
the trade in number of shares.  For each trade we create an artificial
price return $r_i=k\epsilon_iV_i^{\beta}$, where $k$ is a
constant. Then for each block of $M$ trades we compute $r=\sum_{i=1}^M
r_i=k\sum_{i=1}^M \epsilon_iV_i^{\beta}$ and $V=\sum_{i=1}^MV_i$.
Since we are using the real order flow we are incorporating the
correct autocorrelation of the signs $\epsilon_i$ and transaction
sizes $V_i$. Figure~\ref{expRsq}(a) shows $E[r^2|V]$ for different
values of $M$ and $\beta = 0.3$ for the British stock Vodafone in the
period from May 2000 to December 2002, a series which contains
approximately $10^6$ trades.
We see that for small values of $M$ the quantity $E[r^2|V]$ follows
the artificial market impact functional form $E[r^2|V]\sim V^{2\beta}=
V^{0.6}$, but when $M$ is large the relation between $E[r^2|V]$ and
$V$ becomes linear. The value $M=40$ is roughly the average number of
trades in a $15$~minute interval.  We also show error bars computed as
specified by Gabaix et al.  We cannot reject the null hypothesis of a
linear relation between $E[r^2|V]$ and $V$ with $95\%$ confidence,
even though we have a large amount of data, and we know by
construction that $\beta$ is quite different from $1/2$.  We have also
performed tests on other stocks, which give similar results.

One can ask whether it makes a difference that we used a fixed number
of transactions rather than a fixed time interval.  To test this we repeat the
procedure using a fixed time interval of $15$~minutes.
Figure~\ref{expRsq}(b) shows the result.  We see an even clearer
linear relation between $E[r^2|V]$ and $V$ than before, so that the
test once again fails.

Why doesn't this test work?  To gain some understanding of this, we
repeat the same test but shuffle the order of the data, which breaks
the correlation structure.  As shown in Figure~\ref{expRsq}(c), the
result in this case is far from linear even when $M=40$, and the test
easily shows that the market impact does not follow a square root law.
Thus, we see that the problem lies in the autocorrelation structure of
the real data.

In conclusion our numerical simulations show that the linearity test
of $E[r^2|V]$ lacks power to test for a square root market impact with
data containing the correlation structure of real data. In fact, even
a deterministic market impact like $r\sim V^{0.3}$ is consistent with
the relation $E[r^2|V]=a+b~V$ for a sufficiently large number of
trades. Doing this for a fixed time interval rather than a fixed
number of trades time makes this even more evident.  Thus the test of
Gabaix et al. provides no evidence that the average market impact
follows a square root law.

\section{Placing error bars on the average market impact}

While there have been many previous studies of average market impact,
they have not included the statistical analysis needed to assign good
error bars.  In this section we present results about average market
impact at the level of individual ticks.  We show that it does not
generally follow a square root law, and that it varies from market to
market and in some cases from stock to stock in a substantial and
statistically significant way.

Realistic error bars for the average market impact are difficult to
assess due to the fact that volatility is a long-memory process
\cite{Beran94,Granger03}.  That is, its time series has a slowly
decaying power law autocorrelation function that is asymptotically of
the form $\tau^{-\kappa}$, with $\kappa < 1$ so that the integral is
unbounded.  This makes error analysis complicated, since data from the
distant past have a strong effect on data in the present.  Because
volatility is long-memory, the price returns that fall in a given
volume bin $V_a$, which are by definition all of the same sign, are
also long-memory.  This means that the errors in measuring market
impact are much larger than one would expect from intuition based on
an IID hypothesis.

\begin{figure}[ptb]
\begin{center}
\includegraphics[scale=0.35]{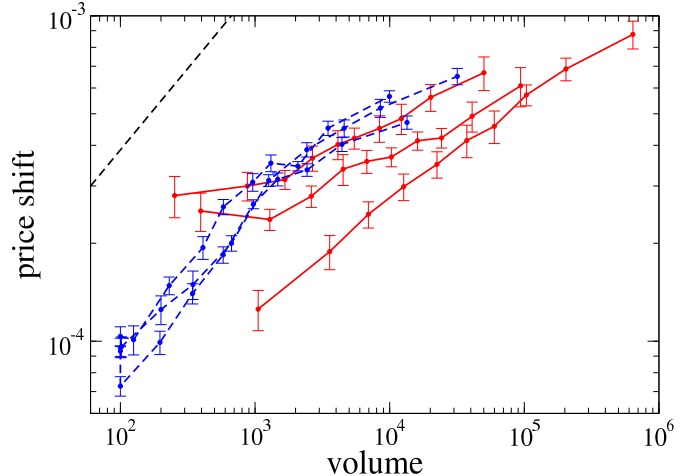}
\caption{Market impact function for buy orders of three stocks 
   traded in the New York Stock Exchange (blue, dashed) and three stocks 
   traded in the London Stock Exchange (red, solid).  Orders of similar
   size $V_i$ are binned together; on the horizontal axis we show the
   average volume of the orders in each bin, and on the vertical axis
   the average size of the logarithmic price change for the orders in
   that bin.  In both cases comparison to the dashed black line in the corner,
   which has slope $1/2$, makes it clear that the behavior for large 
   volume does not follow a law of the form $r_i \sim V_i^{1/2}$.  Error
   bars are computed using the variance plot method \cite{Beran94} as 
   described in the text.}
\label{mktImpactFig}
\end{center}
\end{figure}

We analyze the market impact only for orders (or portions of orders)
that result in immediate transactions.  We call the portion of an
order that results in an immediate transaction an effective market
order, and for the remainder of the paper $V_i$ represents effective
market order size rather than transaction size.  Each order of size
$V_i$ generates a price return $r_i = \log p_a - \log p_b$, where
$p_b$ is the midpoint price quote just before the order is placed and
$p_a$ is the midpoint price quote just after.  We analyze buy and sell
orders separately.  The electronic (SETS) data for the LSE has the
advantage that the data set contains a record of orders, and so we can
distinguish buy and sell orders unambiguously, but has the
disadvantage that it omits trades made in the upstairs
market\footnote{The relative impact on price formation of the upstairs
  and downstairs markets is not clear.  On one hand, the upstairs
  market contains the largest trades.  On the other hand, because
  these trades are arranged privately and then printed in the
  transaction record later, they may not have as large an effect on
  price formation.}.  For the NYSE data we use the trades and quotes
(TAQ) data to infer orders and their signs using the Lee and Ready
algorithm \cite{Lee91}; to identify orders we lump together all trades
with the same timestamp and order code.  To estimate the average
market impact we sort the events $(V_i$, $r_i$) with the same sign
$\epsilon_i$ into bins based on $V_i$ and plot the average value of
$V_i$ for each bin against the average value of $r_i$, as shown in
Figure~\ref{mktImpactFig}.  We choose the bins so that each bin has
roughly the same number of points in it.

To assign error bars for each bin we use the variance plot
method~\cite{Beran94}. For each bin we split the events into $m$
subsamples with $n = K/m$ points, where $K$ is the number of records
in the bin. The subsamples are chosen to be blocks of values adjacent
in time.  For each subsample $i$ we compute the mean $\mu_i^{(n)}$,
$i=1,...,m$.  Then we compute the standard deviation of the
$\mu_i^{(n)}$ which we indicate as $\sigma^{(n)}$. By plotting
$\sigma^{(n)}$ versus $n$ in a log-log plot we compute the Hurst
exponent $H$ by fitting the data with a power-law function
$\sigma^{(n)}=A n^{H-1}$. We compute the error in the mean of the
entire sample of $K$ points by extrapolating the fitted function to
the value $m=K$, i.e. $\sigma=\hat A~K^{\hat H-1}$ where $\hat A$ and
$\hat H$ are the ordinary least square estimate of the parameters $A$
and $H$.  Interestingly, for smaller values of $V_i$ we find 
Hurst exponents substantially larger than $1/2$, whereas for large
values of $V_i$ the Hurst exponents are much closer to $1/2$.
When $H > 1/2$ the error bars are typically much larger than
standard errors\footnote{Since we choose the bins to have
roughly the same number of points, the difference in Hurst exponent
between bins with large and small $V$ cannot be due to a difference
in the mean interval between samples.}.
 
In Figure \ref{mktImpactFig} we show empirical measurements of the
average market impact for the New York Stock Exchange and for the
London Stock Exchange.  We consider three highly capitalized stocks
for each exchange, Lloyds (LLOY), Shell (SHEL) and Vodafone (VOD) for
the LSE, and General Electric (GE), Procter \& Gamble (PG) and AT\&T
(T) for the NYSE. For LSE stocks we consider the period May 2000-
December 2002, while for NYSE stocks we consider the time period
1995-1996. The data for the NYSE are consistent with results reported
earlier without error bars \cite{Lillo03}, while the LSE market impact
data is new. The NYSE data clearly do not follow a power law across
the whole range, consistent with earlier results in references
\cite{Plerou02,Lillo03}.  While $\beta(V_i) \approx 0.5$ for small
$V_i$, for larger $V_i$ it appears that $\beta(V_i) < 0.2$. As shown
in reference \cite{Lillo03}, this transition occurs for smaller values
of $V_i$ for stocks with lower capitalization.  Thus, the assumption
that $\beta = 0.5$ breaks down for high volumes, precisely where it is
necessary in order for the theory of Gabaix et al. to hold.  For the
London data the power law assumption seems more justified across the
whole range, but the exponent is too low; a least squares fit gives
$\beta \approx 0.26$.  While we have not attempted to compute error
bars for the regression, a visual comparison with the error bars of
the individual bins makes it quite clear that $\beta = 1/2$ is
inconsistent with either the London or the NYSE data.  A separate
study of eleven LSE stocks gives $\beta = 0.26 \pm 0.02$ for buy
orders and $0.23 \pm 0.02$ for sell orders \cite{Farmer03}; in as yet
unpublished work this has been extended to $50$ stocks, with similar
results.  Our earlier study for the NYSE was based on 1000 stocks
\cite{Lillo03}.  It is clear that the average market impact functions
are qualitatively different for LSE and NYSE stocks, and that for NYSE
stocks the functional form varies with market capitalization
\cite{Lillo03}.

Even if we abandon the prediction that the average market impact is a
square root law, one might imagine that we could explain fluctuations
in prices in terms of fluctuations in volume modulated by average
market impact of the form $r_i = kV_i^\beta$.  However, if this were
true, for the NYSE the predicted exponent for price fluctuations would
be $\alpha = \gamma/\beta \approx 1.5/0.25 = 6$, which is much too
large to agree with the data. (A typical value \cite{Plerou99} is
$\alpha \approx 3$).  To make matters even worse, the power law
hypothesis for volume or market impact appears to fail in some other
markets.  In the Paris Stock Exchange Bouchaud et
al. \cite{Bouchaud03} have suggested that the average market impact
function\footnote{For the NYSE the logarithmic form for average market
  impact is a reasonable approximation for small $V_i$, but breaks
  down for higher $V_i$} is of the form $\log V_i$, yielding $\beta
\rightarrow 0$ in the limit as $V_i \rightarrow \infty$.  For the
London Stock Exchange the power law hypothesis for average market
impact seems reasonable, but with an exponent significantly smaller
than $1/2$.  Moreover, the volume for the electronic market is not
power law distributed, as discussed in the next section.

Note that we are making all the above statements for individual
orders, whereas many studies have been done based on aggregated data
over a fixed time interval.  Aggregating the data in time complicates
the discussion, since the functional form of the market impact
generally depends on the length of the time interval.  Hence it is
more meaningful to do the analysis based on individual transactions.

\section{Volume distribution}

\begin{figure}[ptb]
\begin{center}
\includegraphics[scale=0.3]{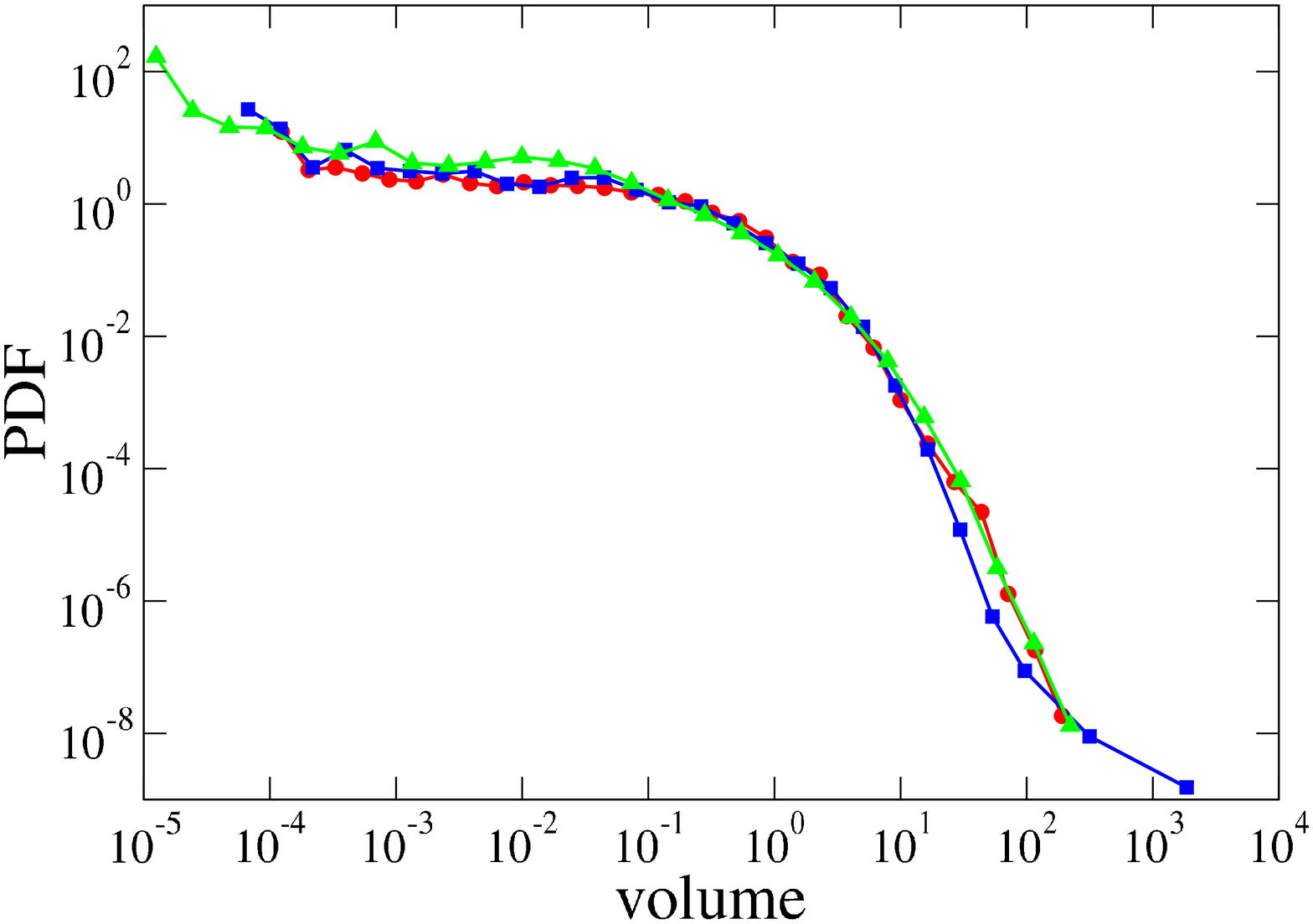}
\includegraphics[scale=0.3]{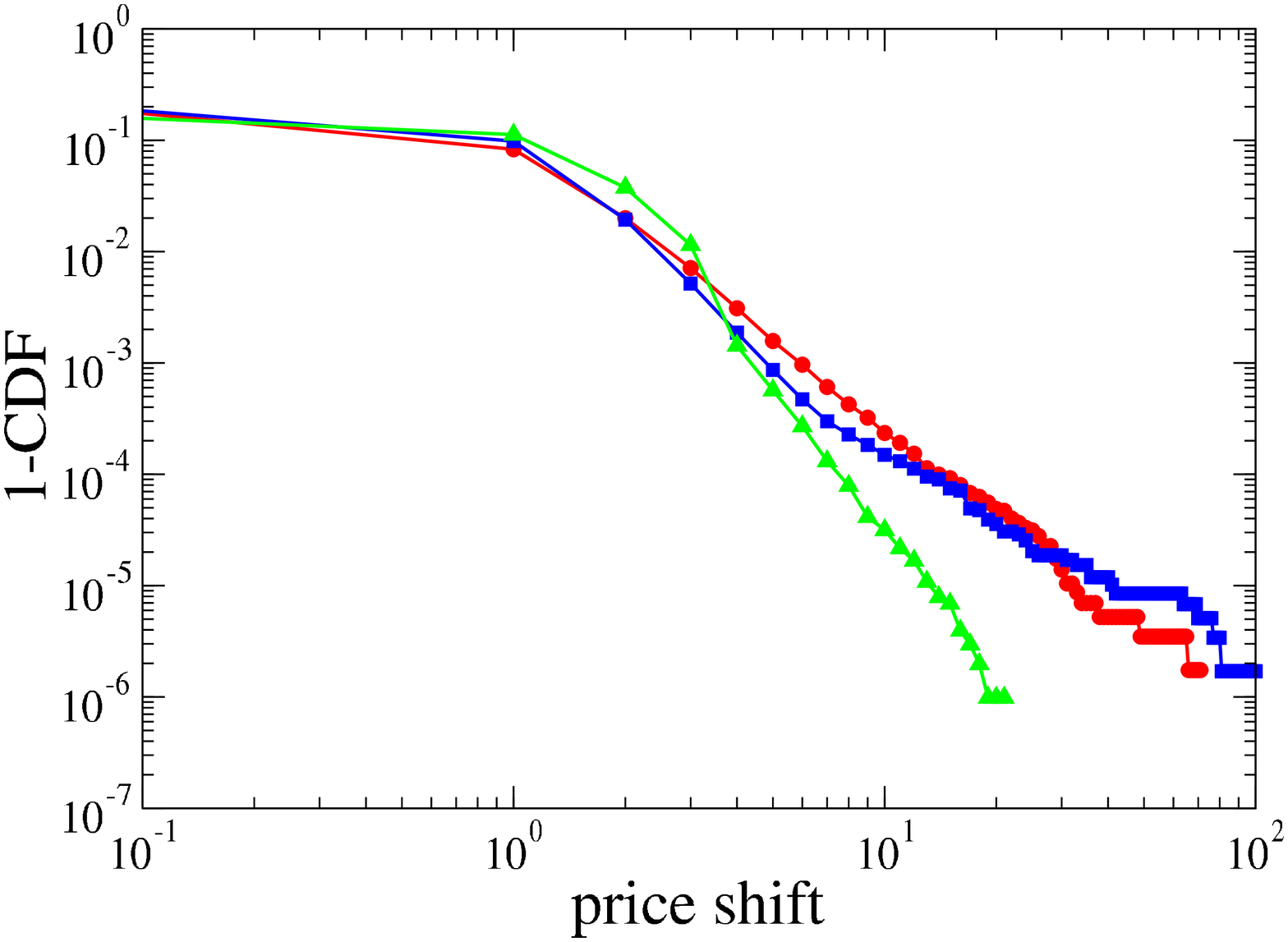}
\caption{(a) The probability density of normalized volume for three
  typical high volume stocks in the LSE, LLOY (red, circles), SHEL
  (blue, squares), and VOD (green, triangles) in the period May 2000-
  December 2002, based on data from the electronic exchange.  There
  are approximately $10^6$ data points for each stock.  (b) $1 -
  P(r)$, where $P(r)$ is the cumulative density function of returns
  induced by the same transactions in (a).  For the normalized volume
  there is no clear evidence for power law tails; in contrast for
  returns this is quite plausible.  Furthermore, the volume
  distributions are essentially identical, whereas the return
  distribution for VOD decays more steeply than the others. }
\label{volDist}
\end{center}
\end{figure}

The theory of Gabaix et al. explains the power law of returns in terms
of the power law of volume, so if volume doesn't have a power law,
then returns shouldn't either.  The existence of a power law tail for
volume seems to vary from market to market.  For the NYSE we
confirm the observation of power law tails for volume reported earlier
\cite{Plerou00}.  However in Figure~\ref{volDist} we show the
distribution of volumes for three stocks in the electronic market of
the LSE.  In order to compare different stocks we normalize the data
by dividing by the sample mean for each stock.
All three stocks have strikingly similar volume distributions; this is
true for the roughly twenty stocks that we have studied.  There is no
clear evidence for power law scaling, even though the power law
scaling of the corresponding return distributions shown in
Figure~\ref{volDist}(b) is rather clear.  If one attempts to fit lines
to the larger volume range of the curve (roughly $10^1 - 10^2$), the
exponent of the cumulative distribution corresponding to
Figure~\ref{volDist}(a) is highly uncertain but it is at least $3$,
which together with the measured values of $\beta$ would imply $\alpha
\approx 3/0.3 \approx 10$.  In contrast, the measured exponents for
Figure~\ref{volDist}(b) are roughly $2.2$, $2.5$, and $4.3$ for SHEL,
LLOY, and VOD respectively.  It is noteworthy that VOD has a much
larger $\alpha$ than the other stocks, even though it has essentially
the same volume distribution and a similar volume distribution; if
anything from Figure~\ref{mktImpactFig} it's $\beta$ is larger than
that of the other stocks, which according to $\alpha = \gamma/\beta$
would imply a smaller $\alpha$.  This provides yet more evidence
that the power law tails of returns are not driven by those of volume.

Note that one of the differences between the NYSE and the LSE data
examined here, which may be the underlying cause of the difference in
their distributions, is that the data from the NYSE includes upstairs
market trades, whereas the LSE data does not.

\section{Conclusion}

We have shown that the conclusions of Gabaix et al. \cite{Gabaix03}
are suspect for three different reasons: First, their statistical
analysis in claiming the existence of a square root law for average
market impact lacks power to reject alternative hypotheses in the
presence of the strong autocorrelations that are present in real data;
Second, new measurements of the average market impact with proper
error bars show that it does not follow a square root law; Third, for
electronic data the London Stock Exchange the distribution of volumes
does not have a power law tail, and there are substantial variations
between the return distributions that are not reflected in variations
in volume or average market impact.  Thus, it seems that the
distribution of large price fluctuations cannot be explained as a
simple transformation of volume fluctuations.

This leaves open the question of what really causes
the power law tails of prices.  We believe that the correct explanation
lies in the extension of theories based on the stochastic properties of
order placement and price formation \cite{Daniels03,Smith03,Farmer03},
which naturally give rise to fluctuations in the response of
prices to orders.  Further work is clearly needed.

Note added in press: In a recent study it has been shown that large
price fluctuations in the NYSE and the electronic portion of the LSE
are driven by fluctuations in liquidity \cite{Farmer03b}.  That is, if
one matches up returns with the orders that generate them, the
conditional distribution of large returns is essentially independent
of order size.  This has been confirmed for the NYSE and Island by
Weber and Rosenow \cite{Weber04}.  The idea that the tail of prices is
driven by fluctuations in liquidity rather than fluctuations in the
number of trades was implicitly suggested earlier by results of Plerou
et al. \cite{Plerou00b}.

\begin{acknowledgments}

We would like to thank the McKinsey Corporation, Credit Suisse First
Boston, The McDonnel Foundation, Bob Maxfield, and Bill Miller for
supporting this research, and Janos Kertesz, Moshe Levy, Rosario
Mantegna, and Ilija Zovko for valuable conversations.

\end{acknowledgments}


\begin{thebibliography}{9}

\bibitem{Gabaix03} Gabaix, X., Gopikrishnan, P., Plerou, V. and Stanley, H.E.
A theory of power-law distributions in financial market fluctuations,
{\it Nature} {\bf 423}, 267-270 (2003).

\bibitem{Plerou02} Plerou, V., Parameswaran, G., Gabaix, X., and 
Stanley, H.E., Qauntifying stock price response to demand
fluctuations, {\it Phys. Rev. E} {\bf 66} 027104 (2002).

\bibitem{Lillo03} Lillo,F. Farmer. J.D. \& Mantegna, R.N., Master
curve for price-impact Function, {\it Nature}
{\bf 421}, 129-130 (2003).

\bibitem{Bouchaud03} Potters, M. \& Bouchaud, J.-P., More statistical 
properties of order books and price impact, {\it Physica A} {\bf 324},
 133-140 (2003)

\bibitem{Bouchaud03b} Bouchaud, J-P., Gefen, Y., Potters, M., and
Wyart, M., Fluctuations and response in financial markets: the subtle
nature of `random' price changes, xxx.lanl.gov/cond-mat/0307332, 2003.

\bibitem{Lillo03b} Lillo, F. and Farmer, J.D., The long memory of the 
efficient market, in preparation.

\bibitem{Beran94} Beran, J. \emph{Statistics for Long-Memory Processes}, Chapman \& Hall (1994).

\bibitem{Granger03} S.-H. Poon and C.W.J. Granger, Forecasting volatility in 
financial markets: A review, {\it J. of Economic Literature} 41, 478-539 (2003).
\bibitem{Lee91}
C. M. C. Lee and M. J. Ready, {\it Journal of Finance}, {\bf 46}
733-746 (1991)

\bibitem{Plerou99} V. Plerou, P. Gopikrishnan, L.A.N. Amaral, M. Meyer 
and H.E. Stanley, Scaling of the Distribution of Price Fluctuations of 
Individual Companies, Phys. Rev. E {\bf 60} 6519-2529 (1999).

\bibitem{Plerou00} P. Gopikrishnan, V. Plerou, X. Gabaix and H.E. Stanley, 
Statistical properties of share volume traded in financial
markets,  Phys. Rev. E {\bf 62} R4493-R4496 (2000).

\bibitem{Daniels03} Daniels, M, Farmer, J.D., Gillemot, L., Iori, G., and Smith,  D.E. Quantitative model of price diffusion and market friction based
on trading as a mechanistic random process, {\it Physical Review
Letters} {\bf 90}, 108102 (2003).

\bibitem{Smith03} Smith, E., Farmer, J.D., Gillemot, L., and 
Krishnamurthy, S., Statistical theory of the continuous double
auction.  To appear in {\it Quantitative Finance}, 2003.

\bibitem{Farmer03} J.D. Farmer, P. Patelli, I.I. Zovko, The predictive power
of zero intelligence in financial markets,
http://xxx.lanl.gov/cond-mat/0309233.

\bibitem{Farmer03b} J.D. Farmer, L. Gillemot, F. Lillo, S. Mike, and
  A. Sen, What really causes large price changes?,
  http://xxx.lanl.gov/cond-mat/0312703 (2003).

\bibitem{Weber04} P. Weber and B. Rosenow, Large stock price changes:
  volume or liquidity?, http://xxx.lanl.gov/cond-mat/0401123 (2004).

\bibitem{Plerou00b} V. Plerou, P. Gopikrishnan, L.A.N. Amaral,
  X. Gabaix, and H.E. Stanley, Economic fluctuations and anomalous
  diffusion, {\it Phys. Rev. E.} {\bf 62} R3023 (2000).

\end{thebibliography}
\end{document}